\newcommand{\HII}{\mbox{H\hspace{0.2em}{\scriptsize II}}}
\newcommand{\Msol}{\mbox{$M_{\sun}$}}
\def\deg{\ensuremath{^\circ}}

\documentclass{aa}
\usepackage{epsfig}
\begin{document}

\thesaurus{05     
           (08.05.1;  
            09.04.1;  
            10.07.2;  
            10.15.2;  
            10.19.2)} 
\title{Cygnus OB2 -- a young globular cluster in the Milky Way}

\author{J.~Kn\"odlseder}
\institute{INTEGRAL Science Data Centre, Chemin d'Ecogia 16, CH-1290 Versoix,
           Switzerland \\
           Centre d'Etude Spatiale des Rayonnements, CNRS/UPS, B.P.~4346,
	       F-31028 Toulouse Cedex 4, France}
\offprints{J\"urgen Kn\"odlseder}
\mail{knodlseder@cesr.fr}
\date{Received 22 March 2000; accepted }
\authorrunning{J. Kn\"odlseder}
\titlerunning{Cygnus OB2 -- a young globular cluster in the Milky Way}
\maketitle

\begin{abstract}

The morphology and stellar content of the Cygnus OB2 association has 
been determined using 2MASS infrared observations in the $J$, $H$, and 
$K$ bands.
The analysis reveals a spherically symmetric association of $\sim2\deg$ 
in diameter with a half light radius of $13'$, corresponding to 
$R_{h} = 6.4$ pc at an assumed distance of 1.7 kpc.
The interstellar extinction for member stars ranges from $A_{V}\approx 
5^{\rm m}$ to $20^{\rm m}$, which led to a considerable 
underestimation of the association size and richness in former optical 
studies.
From the infrared colour-magnitude diagram, the number of OB member 
stars is estimated to $2600 \pm 400$, while the number of O 
stars amounts to $120 \pm 20$.
This is the largest number of O stars ever found in a galactic 
massive star association.
The slope of the initial mass function has been determined from the 
colour-magnitude diagram to $\Gamma=-1.6 \pm 0.1$.
The total mass of Cyg OB2 is estimated to $(4-10)\times 10^4$ \Msol, 
where the primary uncertainty comes from the unknown lower mass 
cut-off.
Using the radial density profile of the association, the central mass 
density is determined to $\rho_{0}=40-150$ \Msol\ pc$^{-3}$.

Considering the mass, density, and size of Cyg OB2 it seems untenable 
to classify this object still as OB association.
Cygnus OB2 more closely resembles a young globular cluster like those 
observed in the Large Magellanic Cloud or in extragalactic star 
forming regions.
It is therefore suggested to re-classify Cygnus OB2 as young globular 
cluster -- an idea which goes back to \cite{reddish66}.
Cygnus OB2 would then be the first object of this class in the Milky 
Way.

\keywords{Stars: early type -- extinction -- globular clusters -- 
open clusters and associations: Cyg OB2 -- Galaxy: stellar content}
\end{abstract}

\section{Introduction}
\label{sec:intro}

The study of galactic OB associations provides the key to a number of 
astrophysical questions, such as 
the star formation process and efficiency,
the interaction of massive stars with the interstellar medium,
the characterisation of the initial mass function at the high-mass 
end,
the study of stellar nucleosynthesis, chemical evolution, and galactic 
recycling processes, and
the evolution of binary systems.
The Cygnus OB2 association is a particularly good region to address 
such questions, since it is extremely rich (e.g.~Reddish et 
al.~1966\nocite{reddish66}, hereafter RLP), and contains some of the most 
luminous stars known in our Galaxy (e.g.~Torres-Dodgen et 
al.~1991\nocite{torres91}).

\begin{figure}[t!]
\epsfxsize=8.5cm \epsfclipon
\epsfbox{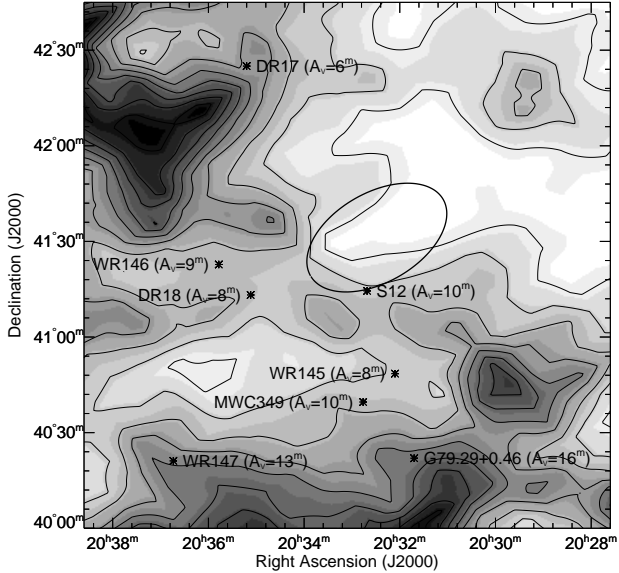}
\caption{Velocity integrated (-10 to 20 km s$^{-1}$) CO intensity map 
of the region around Cyg OB2 (from Leung \& Thaddeus 
1992).
The ellipsoid indicates the size of Cyg OB2 as determined by RLP.
Massive stars that may be associated to Cyg OB2 but lying outside the 
classical Cyg OB2 boundary of RLP are indicated by asterisks.
Visual extinction estimates are quoted in parentheses.}
\label{fig:comap}
\end{figure}

The most comprehensive study of the size and shape of Cyg OB2 has been 
performed by RLP who inferred an elliptical shape with major 
and minor axes of $48'$ and $28'$, respectively (see also 
Fig.~\ref{fig:surface}).
They estimate more than $3000$ members of which at least $300$ 
are of OB spectral type, resulting in a total stellar mass between 
$(0.6-2.7) \times 10^4$ \Msol.
For their analysis, RLP performed star counts on the blue 
and red plates of the Digitized Sky Survey (DSS), reaching limiting 
magnitudes around 20$^{\rm m}$.
Although this limit assures a reasonable complete census for 
unobscured associations, the extreme reddening in and around Cyg OB2 
hampers the detection of even OB stars.
With an estimated distance of 1.7 kpc (e.g.~Massey \& Thompson 
1991\nocite{massey91}) Cyg OB2 is located behind the Great Cygnus 
Rift, causing visual extinction $A_{V}$ from 4$^{\rm m}$ to at least 
10$^{\rm m}$.
A number of observations suggest that Cyg OB2 could indeed be larger than 
the RLP estimate, and that the observed morphology is rather an artifact 
of the particular extinction pattern in the field.
The association boundary determined by RLP fits suspiciously 
well in a region of low CO column density (cf.~Fig.~\ref{fig:comap}) 
and low visual extinction (Dickel \& Wendker 1978\nocite{dickel78}), 
indicating that the visual star densities are probably biased by the 
extinction pattern.
There is a considerable number of early-type stars in the 
obscured area south and south-east of Cyg OB2 that are estimated to lie 
at the same distance as Cyg OB2, and that could indeed be bright member 
stars of the association.
Examples are the Wolf-Rayet stars 
WR 145 and WR 146 (Niemela et al.~1998\nocite{niemela98}),
the potential Luminous Blue Variable star 
G79.29+0.46 (Higgs et al.~1994\nocite{higgs94}),
the massive binary system MWC 349 (Cohen et al.~1985\nocite{cohen85}), 
or the recently discovered group of massive stars around the \HII\ 
region DR 18 (Comer\'on \& Torra 1999\nocite{comeron99}).

The availability of the {\em Two Micron All Sky Survey} (2MASS) 
provides now an excellent opportunity to re-address the question on 
the morphology and stellar content of Cyg OB2.
This survey covers the infrared bands $J$, $H$, and $K$ which have proven to 
be an excellent tool for unveiling embedded star clusters due to the 
reduced impact of dust extinction at longer wavelengths.
In the following I will use these data to determine the morphology 
and stellar content of Cyg OB2.
It will turn out that the association is much larger and much richer
than previously thought, making it the most massive young stellar 
association known in our Galaxy.

\section{Sample selection and colour-magnitude diagrams}
\label{sec:sample}

\begin{figure}[t!]
\epsfxsize=8.5cm \epsfclipon
\epsfbox{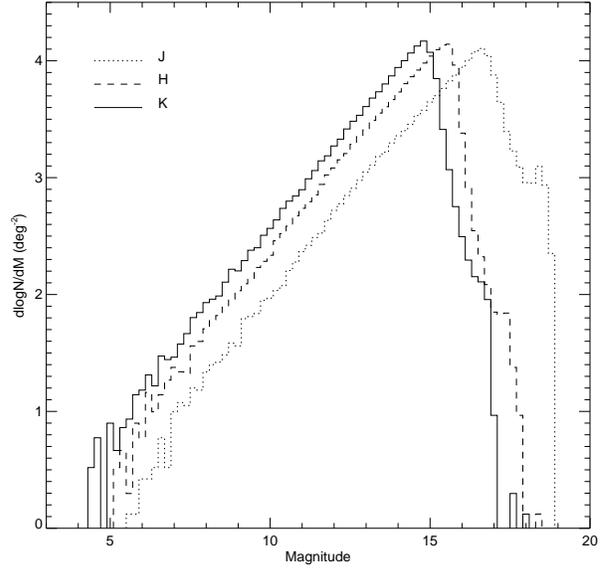}
\caption{Differential $J$, $H$, and $K$ PSC star counts for a square field 
of $2\deg45'$ in size, centred on $\alpha=20^{\rm h}33^{\rm m}6^{\rm s}$ and 
$\delta=+41\deg22'30''$.}
\label{fig:complete}
\end{figure}

The analysis presented in this work is based on the second incremental
release of the 2MASS point source catalogue (PSC) that contains positional 
and photometric data for 162 million objects in the $J$, $H$, and $K$ 
bands.
For the investigation of Cyg OB2, all stars within a square field of 
$2\deg45'$ in size, centred on $\alpha=20^{\rm h}33^{\rm m}6^{\rm s}$ and 
$\delta=+41\deg22.5'$ (J2000)\footnote{All coordinates given in this 
paper are for the epoch J2000.} have been extracted from the 
catalogue, resulting in a total of 203318 selected objects.
Differential star count distributions for these stars in the 3 photometric 
bands are shown in Fig.~\ref{fig:complete}.
The positions of the turn-over in these distributions suggest that the 
selected sample is complete to magnitudes of 16.6, 15.5, and 14.8 in the 
$J$, $H$, and $K$ bands, respectively.
This is 0.5-0.8 magnitudes fainter than the nominal completeness 
limits of the PSC, as expected for less crowded areas in the galactic 
plane (Cutri et al.~2000\nocite{cutri00}).
However, the limits have to be understood as an average for the 
investigated field, and enhanced crowding towards the centre of Cyg 
OB2 may lower the actual completeness magnitude for the association.

\begin{figure*}[th]
\epsfxsize=18cm \epsfclipon
\epsfbox{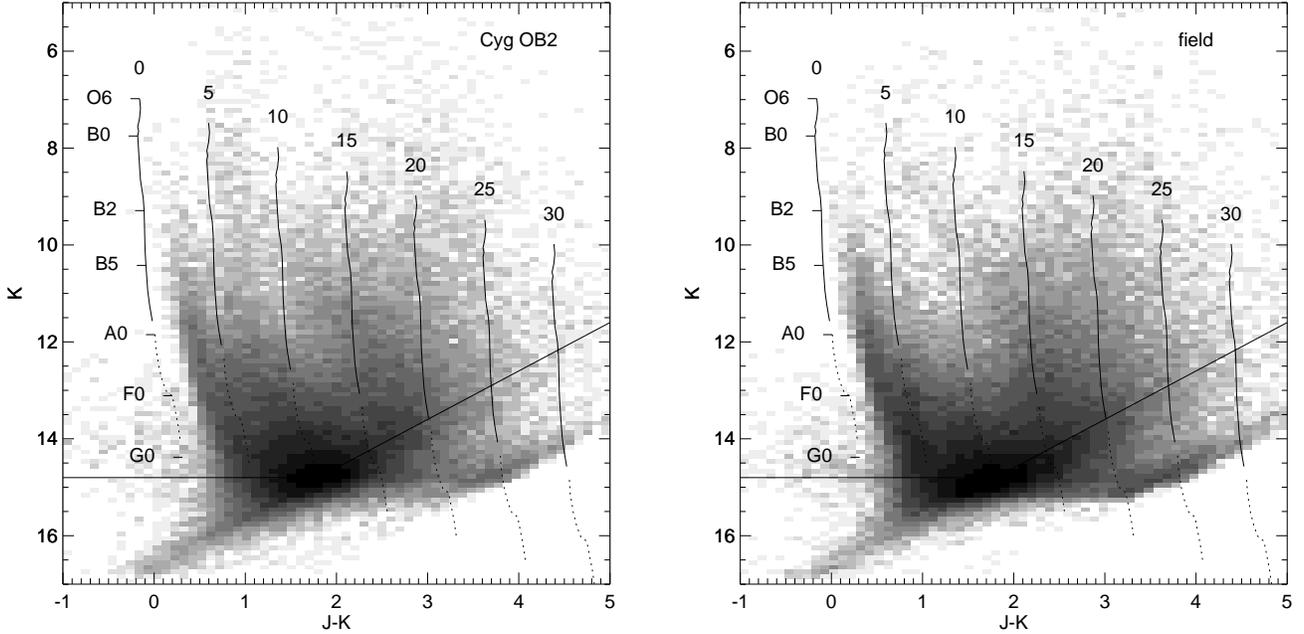}
\caption{Colour magnitude diagram for the Cyg OB2 region (left) and the 
field around Cyg OB2 (right).
The estimated completeness limit is indicated as solid line.
The unreddened main sequence for a distance modulus of 11.2 is shown as 
solid curve for OB stars and as dotted curve for A and F stars.
Reddened main sequences are also shown for $A_{V}=5$ to $30$ in steps 
of 5 magnitudes.}
\label{fig:cmds}
\end{figure*}

\begin{figure}[t!]
\epsfxsize=8.5cm \epsfclipon
\epsfbox{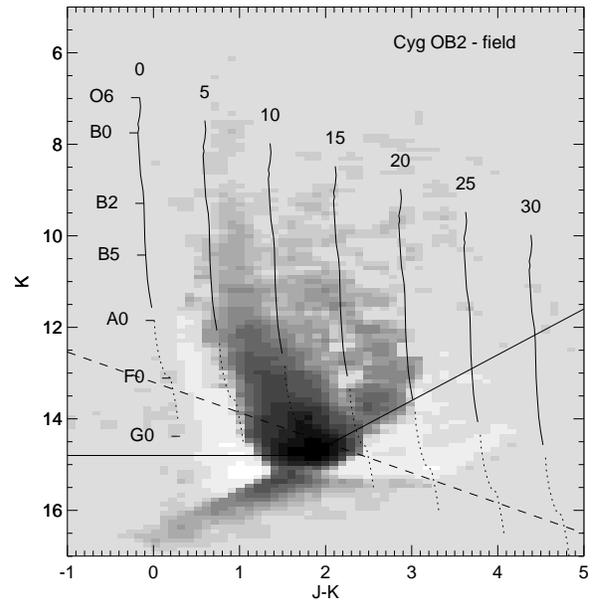}
\caption{CMD of the Cyg OB2 association. The dashed line 
indicates the colour dependent $K$ magnitude selection applied to 
extract the stellar density distribution, limiting the sample to main 
sequence stars of spectral type F3V and earlier.}
\label{fig:cmdnet1}
\end{figure}

\begin{figure*}[th]
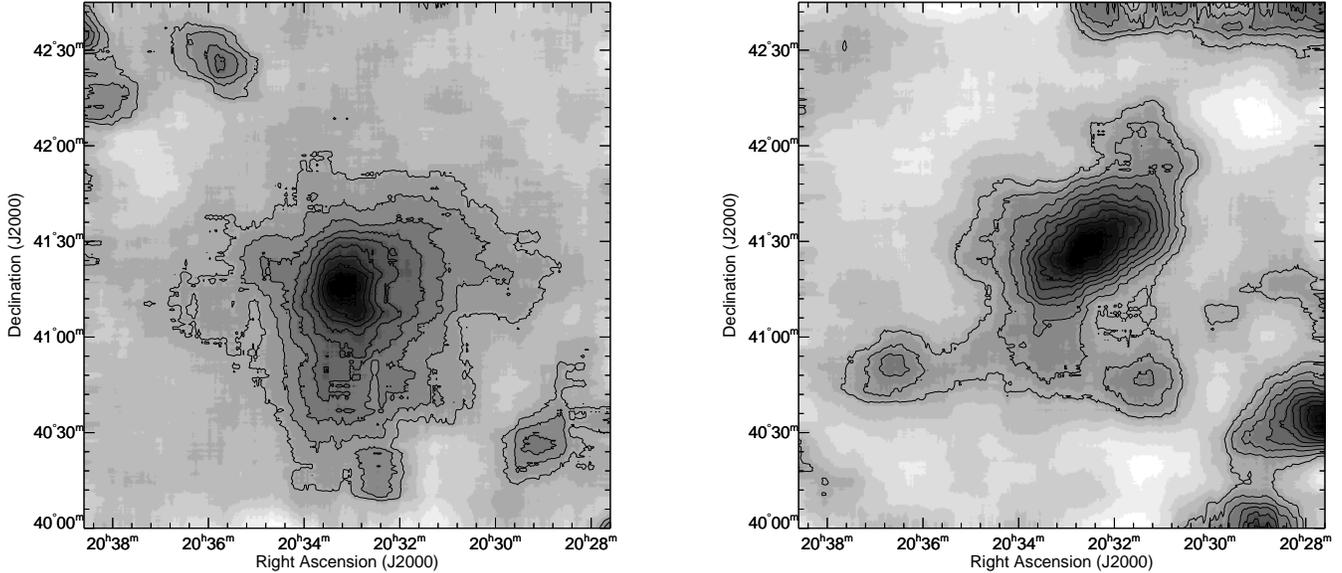

\epsfxsize=8.5cm \epsfclipon
\epsfbox{h2244.f5a}
\hfill
\epsfxsize=8.5cm \epsfclipon
\epsfbox{h2244.f5b}
\caption{Stellar density distributions derived from the 2MASS PSC 
catalogue (left) and DSS red plates (right). 
Dark regions correspond to high densities. 
Isodensity contours are superimposed, starting above the background 
stellar density in 10\% steps with respect to maximum density.
The stellar densities were smoothed with a boxcar average of $15'$ in order 
to emphasise the morphology of the association.}
\label{fig:surface}
\end{figure*}

Colour-magnitude diagrams (CMDs) of the selected stars are presented in 
Fig.~\ref{fig:cmds} as grey-scale plots, which have been obtained by 
counting the number of stars within pixels of 0.1 magnitudes in $J-K$ 
and $K$.
For illustration, the sample has been divided into two distinct 
subsets:
\begin{itemize}
\item the {\em Cyg OB2 sample}, containing all stars within a circular 
      region of $1.05\deg$ in radius, centred on $\alpha=20^{\rm 
      h}33^{\rm m}10^{\rm s}$ and $\delta=+41\deg 12'$
\item the {\em field star sample}, containing all remaining stars outside 
      the circular region.
\end{itemize}
The choice of the separation was motivated by the location and size of 
the Cyg OB2 association (cf.~Sect.~\ref{sec:content}).
Geometrically, the Cyg OB2 sample covers a field of 3.46 degrees 
squared while the field star sample covers 4.10 degrees squared.
The number of stars in the CMDs is 100754 for the Cyg OB2 and 
102167 for the field stars sample; 397 stars lie outside the plotted 
range.
Normalising the number of stars in the field star sample to the 
geometrical area of the Cyg OB2 sample and subtraction from the 
Cyg OB2 sample results in an excess of 14420 stars within the 
association field.
The CMD of these excess stars is shown in Fig.~\ref{fig:cmdnet1}.
This figure has been obtained by subtracting the field star CMD, multiplied 
by 0.85 to account for the different geometrical areas, from the Cyg OB2 
sample CMD.
To reduce statistical fluctuations the resulting CMD was smoothed with a 
boxcar average of 0.3 magnitudes in $J-K$ and $K$.

Figure \ref{fig:cmdnet1} clearly illustrates the wide spread of visual 
extinctions in Cyg OB2, covering the range from $A_{V} \approx 5$ to 
$20$ magnitudes.
Between $A_{V} \approx 5-10$ magnitudes, a clear main sequence 
distribution is visible which spatially coincides with regions of low 
visual obscuration, as depicted by the light areas in the CO map 
(cf.~Fig.~\ref{fig:comap}).
Stars more heavily reddened are related to regions of higher CO 
column density, located in particular south of the classical Cyg OB2 
boundaries.

In the CMDs the PSC sample completeness limits manifest as a colour 
dependent $K$ magnitude limit.
For $J-K < 1.8$, the $K$ magnitude limit of 14.8 is driving the sample 
completeness, while for $J-K > 1.8$ the $J$ magnitude limit of 16.6
becomes dominant, resulting in a colour dependent completeness limit of 
$K = 16.6 - (J-K)$.
The resulting limit is shown in Figs.~\ref{fig:cmds} and 
\ref{fig:cmdnet1} as solid line.
For illustration, the unreddened main sequence for O to F type stars 
with a distance modulus of 11.2, as suggested by \cite{massey91} for 
Cyg OB2, is also shown (see Appendix \ref{sec:calibrations} for a 
reference of the photometric calibrations).
Reddened main sequences are also shown for $A_{V}=5$ to $30$ in steps 
of 5 magnitudes, assuming $A_{V}/A_{K}=10$ and a reddening slope of
$A_{K}/E(J-K)$ of 0.66 (Rieke \& Lebofsky 1985\nocite{rieke85}).

Comparison of the reddened main sequences with the completeness limit 
suggests that the PSC is complete for OB stars in Cyg OB2 up to 
visual extinctions of $A_{V} \approx 20$ magnitudes.
This is comparable to the maximum reddening found for the association 
stars (cf.~Fig.~\ref{fig:cmdnet1}), hence the OB stars sample in the 
PSC should be complete.

\section{Morphology from star counts}
\label{sec:starcounts}

To study the morphology of the Cyg OB2 association, the stellar 
density distribution of the field has been determined by binning the 
PSC stars in pixels of size $30'' \times 30''$.
Only stars with $K$ magnitude brighter than $13.2 + 0.66 \times (J-K)$ 
have been selected in order to minimise any bias through variable 
extinction, and to maximise at the same time the number of stars in 
the stellar density distribution.
This selection is equivalent to choosing stars intrinsically brighter 
than $M_{K} \le 2.2^{\rm m}$ at the distance of Cyg OB2, corresponding 
to main sequence stars of spectral type F3V and earlier (cf.~Appendix 
\ref{sec:calibrations}).
A comparison to the completeness limit in Fig.~\ref{fig:cmdnet1} shows 
that stars later than spectral type B9 and suffering from visual extinction 
in excess of $A_{V} \approx 12^{\rm m}$ could be missed in the sample, but 
the CMD of Cyg OB2 suggests that only few association stars show these 
properties.
Hence, the stellar density distribution should be relatively unaffected 
by absorption.

The presence of several bright stars in the field results in a couple 
of blind areas in the PSC where no information is available.
These blind areas were filled by estimating their stellar densities from 
adjacent fields.
A small ($\sim10 \%$) gradient in the field star density distribution 
has been removed by subtracting a 2-dimensional linear function that was 
estimated by fitting a lower envelope to the data.
The resulting stellar density distribution is shown in the left panel of 
Fig.~\ref{fig:surface}.
For display, the distribution has been smoothed with a boxcar average 
of $15'$ to reduce statistical density fluctuations and to emphasise the 
global morphology of the association.
Still, most of the fine structure in the image is probably of 
statistical nature and should be interpreted with care.

The image reveals a rather regular and almost circular stellar density 
profile for Cyg OB2 with a pronounced maximum at 
$\alpha=20^{\rm h}33^{\rm m}10^{\rm s}$ and $\delta=+41\deg 15.7'$.
The maximum is slightly offset from the centre, which is determined to
$\alpha=20^{\rm h}33^{\rm m}10^{\rm s}$ and $\delta=+41\deg 12'$ 
(see Sect.~\ref{sec:content}).
Star counts in concentric radial annuli around the centre show that 
$50\%$ of the members are located within a radius of $21'$, and $90\%$ 
within a radius of $45'$ around the centre.
The association merges with the field stars at a radius of $\sim1\deg$.
Assuming a total diameter of $2\deg$ results in a physical diameter of 
60 pc at a distance of 1.7 kpc.

A slight density enhancement at $\alpha=20^{\rm h}32^{\rm m}12^{\rm s}$ 
and $\delta=+40\deg 21'$ at the southern edge of Cyg OB2 possibly 
indicates some substructure in the association.
The stars in this area show $J-K$ colours $>2^{\rm m}$, 
and are compatible with heavily reddened ($A_{V}>12^{\rm m}$) main sequence 
objects.
The feature coincides spatially with the \HII\ region DR 15 which has 
been identified as a nursery of embedded B stars (Odenwald et 
al.~1990\nocite{odenwald90}).
However, the distance of DR 15 is quite uncertain ($1-4.2$ kpc; 
Wendker et al.~1991\nocite{wendker91}), making the physical association 
to Cyg OB2 highly speculative.

Two other features of localised stellar density enhancements are 
visible in Fig.~\ref{fig:surface}.
At $\alpha=20^{\rm h}35^{\rm m}30^{\rm s}$ and $\delta=+42\deg 25'$ a 
group of red ($J-K > 2^{\rm m}$) stars coincide spatially with the 
extended \HII\ region DR 17 (Wendker et al.~1991\nocite{wendker91}).
Comparison with the CO map (cf.~Fig.~\ref{fig:comap}) suggests that 
also this feature could be an embedded star nursery, but its angular 
separation and estimated distance of 800 pc - 1.5 kpc (Odenwald \& 
Schwartz 1993\nocite{odenwald93}) make a physical relation to Cyg OB2 
highly improbable.
Another group of stars is found at $\alpha=20^{\rm h}29^{\rm m}5^{\rm s}$ 
and $\delta=+40\deg 25'$ which spans a wide range of colours, and 
which does not coincide with any known \HII\ region or stellar cluster.
The angular separation from Cyg OB2 suggests that this group is also 
not related to the association.

The morphology of Cyg OB2 as derived from the 2MASS catalogue is 
quite different to that inferred by RLP from DSS plates.
While RLP found an elliptically shaped association of $29\times17$ pc 
in size, the PSC analysis presented in this work suggests a spherical 
association of 60 pc in diameter.
To investigate the origin of this discrepancy, the stellar density 
distribution around Cyg OB2 has also been derived from the DSS red plates,
reproducing Fig.~23 of RLP.
The result of this analysis is shown in the right panel of 
Fig.~\ref{fig:surface}, again smoothed by a boxcar average of $15'$ 
to enhance the morphology of the association.
Indeed, the 2MASS and the DSS stellar density distributions are 
markedly different.
Comparison with the matter distribution in the region 
(cf.~Fig.~\ref{fig:comap}) clearly shows that the morphology of the DSS 
density distribution is largely influenced by the absorption pattern.
The elliptical shape of Cyg OB2 in the DSS data is produced by an elongated 
region of rather low extinction that is enclosed by two dust lanes running 
from south-east to north-west.
Several local maxima that appear like extensions from the Cyg OB2 
association are simply areas of reduced absorption, where the number 
of background stars is locally enhanced.
A massive system of molecular clouds below $\delta=+40\deg 30'$ 
provides an efficient barrier for visible light, pushing the apparent 
centroid of Cyg OB2 towards the north.
Hence DSS star counts are apparently not very suitable for the 
determination of the morphology of Cyg OB2.
In contrast, a comparison with Fig.~\ref{fig:comap}, and the above
discussion of the completeness of the PSC sample, suggest that the 
2MASS stellar density distribution is basically unaffected by absorption, 
hence it reveals indeed the true morphology of Cygnus OB2.

\section{Size, mass, and stellar content}
\label{sec:content}

\begin{figure}[t!]
\epsfxsize=8.5cm \epsfclipon
\epsfbox{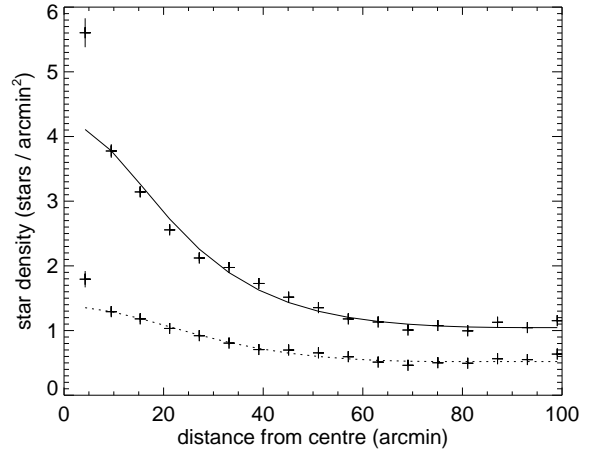}
\caption{Radial star density distribution for stars earlier than 
spectral type F3V (solid) and only OB type stars (dotted).
The crosses present the observed stellar densities while the lines are
fitted King profiles.
An additional background star density was subtracted prior to the analysis 
in order to remove a small gradient ($\sim10\%$) in the field star 
density.
For the F3V sample, the zero point corresponds to a field star density 
of 3.1 stars arcmin$^{-2}$ while for the OB star sample the zero point 
amounts to 0.8 stars arcmin$^{-2}$.}
\label{fig:radial}
\end{figure}

In order to determine the size, centre, and stellar content of Cyg OB2, 
radial star density profiles have been extracted from the 2MASS data by 
counting the number of stars intrinsically brighter than 
$M_{K} \le 2.2^{\rm m}$ in concentric radial annuli around an assumed 
centre, divided by the annuli surface.
Geometrical corrections have been applied for the outer annuli that 
partially fall outside the rectangular survey region.
The profiles were fitted by a King law (King 1962\nocite{king62})
\begin{equation}
  f(r) = k \{ [ 1+(r/r_{\rm c})^2 ]^{-1/2} - 
              [ 1+(r_{\rm t}/r_{\rm c})^2 ]^{-1/2} \}^2
\end{equation}
on top of a constant to determine the core radius $r_{\rm c}$, the tidal 
radius $r_{\rm t}$, and the field star density.
By searching the central position that minimises the radial extent of the 
profile the centre of Cyg OB2 has been determined to
$\alpha=20^{\rm h}33^{\rm m}10^{\rm s}$ and $\delta=+41\deg 12'$.
The corresponding density profile is shown in Fig.~\ref{fig:radial}.
The central stellar density reaches $4.5$ stars arcmin$^{-2}$ above 
the field star density, and drops to $50\%$ at a radius of $13'$, 
resulting in a half light radius of $R_{h}=6.4$ pc at a distance of 
1.7 kpc.

Best fitting King parameters for the profile are $r_{c}=29'\pm5'$ and 
$r_{t}=93'\pm20'$, leading to a concentration parameter
$\log r_{t}/r_{c}$ of $0.5$.
The reduced $\chi^2$ of the fit is only $10.9$, indicating that the 
King law is not a very accurate description of the radial density 
profile.
Indeed, there is no physical reason to believe that Cyg OB2 should 
follow a King profile.
The basic aim of using King profiles was the estimation of the 
field star density, and comparison with Fig.~\ref{fig:radial} 
convinces that at least this goal was reached.

By integrating the radial profile after subtraction of the fitted 
field star density, the total number of association stars with 
$M_{K} \le 2.2^{\rm m}$ (corresponding to spectral types of F3V and earlier) 
amounts to $8600 \pm 1300$.
The error (as all following error quotations) includes a possible systematic 
uncertainty from the field star subtraction.
This systematic uncertainty was estimated by applying different methods for 
the extraction of the association stars, such as analysing star density 
profiles along constant Right Ascension or declination, estimating the
field star density from a circular region around the association, or 
by performing the analysis without removing the gradient in the field 
star density distribution.

Selecting only $K$ magnitudes brighter than $11.55 + 0.66 \times 
(J-K)$ limits the sample to stars intrinsically brighter than
$M_{K} \le 0.37^{\rm m}$, corresponding roughly to spectral type B9V 
and earlier.
Repeating the radial density profile analysis for this limited sample 
allows the determination of the total number of OB stars in Cyg OB2 
to $2600 \pm 400$.
This number is at the upper end of the range quoted by RLP (300 - 
3000), confirming their suggestion that many of the highly reddened 
stars in the DSS survey are indeed OB stars.
Further, restricting $M_{K} \le -3.55^{\rm m}$, by requiring 
$K \le 7.76 + 0.66 \times (J-K)$, selects only O stars from 
the PSC, resulting in a total number of $120 \pm 20$ objects.
In their survey, \cite{massey91} find 40 O stars within a central 
field of 0.35 degrees squared of the Cyg OB2 association.
Integrating over the same region in the 2MASS data gives $39 \pm 6$ O 
stars, demonstrating that the present analysis is in excellent agreement 
with the Massey \& Thompson\nocite{massey91} observations.
Hence, the large number of O stars found in this analysis is mainly 
due to the large extent of the association, which previously was 
missed due to the high extinction in the area.

Taking an initial mass of 1.5 \Msol\ for a F3V star (Schmidt-Kaler 
1982\nocite{schmidt82}), and assuming the slope $\Gamma$ of the initial 
mass function (IMF)\footnote{The 
Salpeter IMF has a slope of $\Gamma=-1.35$ in this prescription.}
to be comprised between $-1.1$ and $-1.7$, the 
number of association stars converts into a total stellar mass of 
$(3-5) \times 10^4$ \Msol\ above 1.5 \Msol.
Extrapolation of the IMF to lower masses using the prescription of 
\cite{kroupa93} results in a total association mass of 
$(4-10) \times 10^4$ \Msol, where the boundaries correspond to a lower 
mass cut-off of $1.0$ and $0.08$ \Msol, respectively.
Since the actual value of the mass cut-off in Cyg OB2 is unknown, the 
total association mass is uncertain to within the quoted limits.
However, even at the lower mass limit, Cyg OB2 would be the most 
massive OB association known in the Galaxy, being comparable 
in mass to a small globular cluster.

From the radial density profile and the total mass estimates, the 
central mass density of Cyg OB2 can be estimated.
Assuming a total mass of $4 \times 10^4$ \Msol\ results in a mass 
density of $\rho_{0} = 40 - 60$ \Msol\ pc$^{-3}$, where the lower 
value comes from the extrapolation of the best fitting King profile 
into the centre, and the upper value comes from the observed central 
stellar density.
Correspondingly, the central mass density estimate for a total mass of
$1 \times 10^5$ \Msol\ amounts to $\rho_{0} = 100 - 150$ \Msol\ 
pc$^{-3}$.
As before, the distance to Cyg OB2 has been assumed to 1.7 kpc.

\section{Initial mass function}
\label{sec:imf}

To derive the mass spectrum of Cyg OB2, sub-samples of the PSC data 
containing all member stars within a given interval 
$[M_{K}^{\rm low}, M_{K}^{\rm up}]$ of intrinsic $K$ magnitudes have 
been defined.
Additionally, the $K$ band extinctions $A_{K}$ were restricted to the 
range $0 \le A_{K} \le 2$ that has been found to enclose the
association members (cf.~Sect.~\ref{sec:sample}).
To define the intrinsic $K$ magnitude intervals, photometric cuts 
along the reddening lines have been applied using
\begin{equation}
 \tilde{K}_{\rm low} \ge K - DM - 0.66 \times(J-K) \ge \tilde{K}_{\rm up}
 \label{eq:cutK}
\end{equation}
where $M_{K} = 0.0037 + 1.036 \times \tilde{K}$.
The extinction cut was realised using the relation
\begin{equation}
 A_{K} \approx \frac{K + 0.057 - 17.835 \times (J-K) - DM}
                    {1 - 17.835 / 0.66}
 \label{eq:Ak}
\end{equation}
which is an approximation of the extinction at the distance of Cyg OB2 as 
function of the position in the CMD for $M_{K} \le 2^{\rm m}$
(see Appendix \ref{sec:photocut} for details about the derivation of 
these relations).
In Eqs.~(\ref{eq:cutK}) and (\ref{eq:Ak}), $DM=11.2$ is the assumed 
distance modulus of Cyg OB2.

\begin{figure}[t!]
\epsfxsize=8.5cm \epsfclipon
\epsfbox{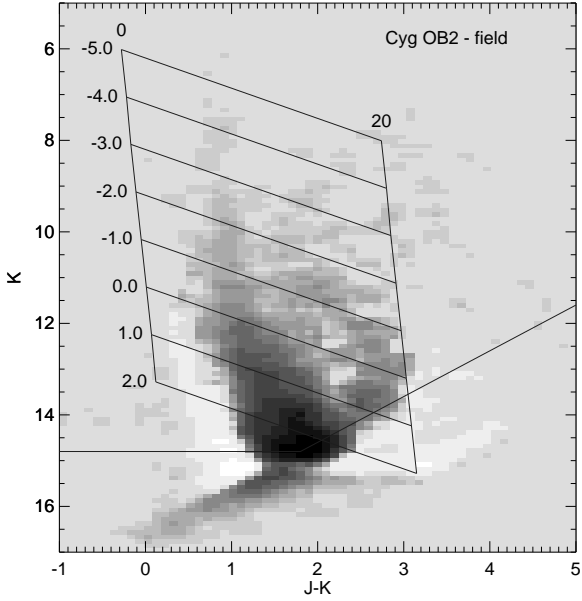}
\caption{CMD of the Cyg OB2 association with photometric cuts 
         superimposed that have been applied to determine the IMF. 
         The limits $\tilde{K}$ of Eq.~(\ref{eq:cutK}) are 
         quoted at the left edge of the selected areas.}
\label{fig:cmdnet2}
\end{figure}

Using these equations, the $K$,$J-K$ plane has been divided into 7 
areas of identical size which define 7 stellar sub-samples 
in the PSC data (cf.~Fig.~\ref{fig:cmdnet2}).
The corresponding $[\tilde{K}_{\rm low}, \tilde{K}_{\rm up}]$ and 
$[M_{K}^{\rm low}, M_{K}^{\rm up}]$ intervals are summarised in Table 
\ref{tab:cuts}.
The $[M_{K}^{\rm low}, M_{K}^{\rm up}]$ intervals have then been 
converted into initial mass intervals using the mass-luminosity relation
derived in Appendix \ref{sec:calibrations}.
The corresponding mass boundaries are quoted in the last two columns of 
Table \ref{tab:cuts}.

\begin{table}[th]
  \footnotesize
  \caption{\label{tab:cuts}
  Selection parameters of the seven sub-samples.}
  \begin{flushleft}
    \begin{tabular}{llllll}
    \hline
    \noalign{\smallskip}
    $\tilde{K}_{\rm low}$ & $\tilde{K}_{\rm up}$ & 
    $M_{K}^{\rm low}$ & $M_{K}^{\rm up}$ & 
    $\log M^{\rm low}$ & $\log M^{\rm up}$ \\
    \noalign{\smallskip}
    \hline
    \noalign{\smallskip}
     2 &  1 &  2.08 &  1.04 & 0.196 & 0.407 \\
     1 &  0 &  1.04 &  0.00 & 0.407 & 0.613 \\
     0 & -1 &  0.00 & -1.03 & 0.613 & 0.826 \\
    -1 & -2 & -1.03 & -2.07 & 0.826 & 1.048 \\
    -2 & -3 & -2.07 & -3.10 & 1.048 & 1.285 \\
    -3 & -4 & -3.10 & -4.14 & 1.285 & 1.654 \\
    -4 & -5 & -4.14 & -5.18 & 1.654 & 2.211 \\
    \noalign{\smallskip}
    \hline
    \end{tabular}
  \end{flushleft}
\end{table}

Two methods have been employed to extract the IMF from the PSC data.
First, radial density profiles have been derived for the 7 sub-samples 
which were then fitted by King profiles in order to estimate the field 
star density in each of the samples.
The field star density estimate was then subtracted from the radial 
density profile, and integration over the residual profile provided 
an estimate of the number of member stars in the sub-sample.
Note that this method properly accounts for possible variations in 
the size of the association for different intrinsic magnitude 
intervals, which could result from mass segregation.
However, the analysis revealed similar radial extents for all 7 
sub-samples, indicating that no mass segregation has taken place in 
Cyg OB2.
The second method consisted simply in integrating the number of stars 
in the field star subtracted CMD in the 7 areas indicated in 
Fig.~\ref{fig:cmdnet2}.
Here, the field star density is estimated from the area beyond 
$1.05\deg$ from the association centre, as discussed in Sect.~\ref{sec:sample}.
Note that in this case the gradient in the field stars density distribution 
has not been removed.

\begin{figure}[t!]
\epsfxsize=8.5cm \epsfclipon
\epsfbox{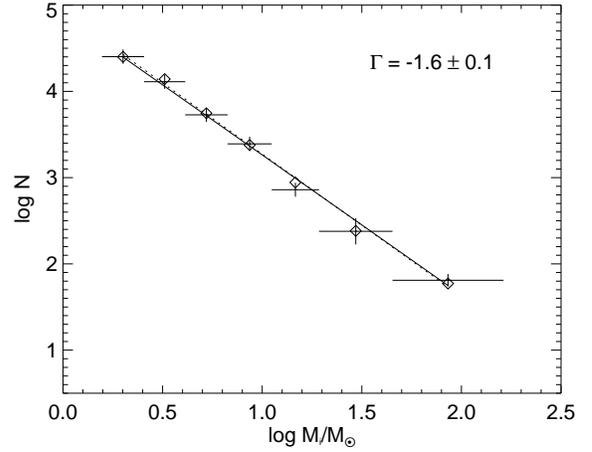}
\caption{IMF for Cygnus OB2.
         The solid line present the best fitting power law to the 
         points derived from radial density profiles (crosses) while
         the dotted line is the best fitting power law to the CMD 
         integrated data (diamonds).}
\label{fig:imf}
\end{figure}

The results of both analyses are summarised in Fig.~\ref{fig:imf}.
The points with error bars were derived using radial density profiles 
(i.e.~the first method), while the results from integrating the CMD are 
show as diamonds.
Apparently, both methods lead to consistent IMFs.
The data points are reasonably well fitted by a power law IMF of slope 
$\Gamma=-1.6\pm0.1$ for both methods.
The slope is compatible with the \cite{kroupa93} analysis of the solar 
neighbourhood, but considerably steeper than $\Gamma=-1.0\pm0.1$ determined 
by \cite{massey91} for the central part of Cyg OB2.
Note, however, that photometric mass determinations, as performed in 
this work, are less reliable than a spectroscopic study, as done by
\cite{massey91}, and a change in the employed mass-luminosity relation 
could easily flatten the IMF.
On the other hand, the \cite{massey91} analysis did not extend to such
low masses as the present analysis does, and the flat slope they 
determined could also reflect an incompleteness in the lower mass 
domain.

Finally, Fig.~\ref{fig:cmdnet2} suggests that the lowest mass bin 
could be subject to incompleteness since the integration region 
extends below the completeness limit.
In particular, the completeness limit for the central part of 
Cyg OB2 may lie at even smaller magnitudes due to enhanced crowding, 
making $\Gamma=-1.6$ formally to a lower limit for the slope of the 
IMF.
Nevertheless, the lowest mass bins in Fig.~\ref{fig:imf} fall not 
particularly below the IMF extrapolation from the higher masses, 
indicating that there are probably not many member stars which are 
missed.
The number of $8600 \pm 1300$ members earlier than spectral type F3V, 
derived in Sect.~\ref{sec:content}, is therefore probably a solid 
value.

\section{Conclusions}
\label{sec:conclusions}

The analysis of 2MASS data from the Cygnus region has revealed that 
Cygnus OB2 is considerably larger and more massive than previously 
thought.
Previous surveys in the visible wavelength range, such as the 
extensive work of RLP, were substantially biased by the local 
extinction pattern in the field, which is caused by a foreground 
molecular cloud structure, known as the Great Cygnus Rift.
The infrared data reveal now a spherically shaped association, with 
stellar extinctions reaching $A_{V} \approx 20^{\rm m}$ in the 
southern part of the field.
Several highly reddened early-type objects that formerly were situated 
outside the association boundary are now lying within the association, 
indicating that they possibly are members of Cyg OB2. 
Examples are the massive binary system MWC349 
(Cohen et al.~1985\nocite{cohen85}),
the potential Luminous Blue Variable star G79.29+0.46 (Higgs et 
al.~1994\nocite{higgs94}), the Wolf-Rayet stars WR 145 and WR 146 
(Niemela et al.~1998\nocite{niemela98}), and the recently discovered group 
of massive stars around the \HII\ region DR 18 
(Comer\'on \& Torra 1999\nocite{comeron99}).
Deep spectroscopic surveys of the stars in the highly obscured regions 
should help to consolidate the stellar population of Cyg OB2, and 
will improve our knowledge about the evolutionary state of the 
association.

\begin{table}[th]
  \footnotesize
  \caption{\label{tab:parameters}
  Summary of Cyg OB2 properties (radial dimensions and the central 
  mass density have been calculated for an assumed distance of 1.7 kpc).}
  \begin{flushleft}
    \begin{tabular}{ll}
    \hline
    \noalign{\smallskip}
    Centre (J2000) & $\alpha=20^{\rm h}33^{\rm m}10^{\rm s}$, 
                     $\delta=+41\deg 12'$ \\
    Diameter                        & $\sim2\deg$ ($\sim$60 pc) \\
    Half light radius $R_{h}$       & $13'$       (6.4 pc) \\
    Core radius $r_{c}$             & $29'\pm5'$  ($14 \pm 2$ pc) \\
    Tidal radius $r_{t}$            & $93'\pm20'$ ($46 \pm 10$ pc) \\
    Members earlier F3V             & $8600 \pm 1300$ \\
    OB star members                 & $2600 \pm  400$ \\
    O star members                  &  $120 \pm   20$ \\
    Total stellar mass              & $(4-10) \times 10^4$ \Msol\ \\
    Central mass density $\rho_{0}$ & $40 - 150$ \Msol\ pc$^{-3}$ \\
    IMF slope $\Gamma$              & $-1.6 \pm 0.1$ \\
    \noalign{\smallskip}
    \hline
    \end{tabular}
  \end{flushleft}
\end{table}

RLP already recognised from their analysis that Cyg OB2 is unusually 
massive and compact, and hence suggested that it should be regarded as 
a new globular cluster of stars, similar to the blue globular clusters 
discovered by \cite{hodge61} in the Large Magellanic Cloud (LMC).
Indeed, the comparison of the association parameters, summarised in 
Table \ref{tab:parameters}, to typical parameters for galactic OB 
associations, young open clusters, and globular clusters leads 
inevitably to this conclusion.
For an OB association, Cyg OB2 is simply too compact.
The typical mass density of OB associations is well below
$0.1$ \Msol\ pc$^{-3}$ (Blaauw 1964\nocite{blaauw64}) and the 
mean size amounts to 137 pc (Garmany 1994\nocite{garmany94}).
Cyg OB2, however, shows an average mass density of $5-12$ \Msol\ 
pc$^{-3}$ within the inner 10 pc and a diameter of only 60 pc.
Hence, in contrast to classical OB association, Cyg OB2 should be a 
gravitationally bound system (Blaauw 1964\nocite{blaauw64}).
\cite{garmany92} suggest that Cyg OB2 might be the nucleus of an OB 
association, hence it could be considered as an open cluster.
However, for an open cluster (as well as for an OB association) Cyg 
OB2 is by far too massive.
Typical masses for open clusters are at most $10^3$ \Msol\ and do not 
exceed $10^4$ \Msol\ (Bruch \& Sanders 1983\nocite{bruch83}) while 
the total mass of Cyg OB2 amounts to $4-10\times10^4$ \Msol.
Such a high mass is more representative for a small globular cluster.

In particular, the parameters of Cyg OB2 (as listed in Table 
\ref{tab:parameters}) correspond fairly well to the typical 
parameters derived for young globular clusters in the LMC
(e.g.~Elson et al.~1987\nocite{elson87}; 
Fischer et al.~1992\nocite{fischer92}).
They have masses $\sim10^4 - 10^5$ \Msol, central densities 
$\sim10^2$ \Msol\ pc$^{-3}$, and extend to radii $\sim80$ pc.
Little or no mass segregation is observed in these objects and they 
show only a small age spread
(Elson et al.~1989\nocite{elson89}).
The same is true for Cyg OB2.
The determination of radial profiles for different intrinsic 
luminosities did not reveal any mass segregation, and the analysis 
of a sub-sample of massive stars in Cyg OB2 by \cite{massey91} did 
not suggest a considerable age spread.
Thus I believe it is reasonable to conclude that Cygnus OB2 shows 
the same properties as the system of young globular clusters in the 
LMC -- hence Cygnus OB2 should be considered as a member of this 
population.

Cyg OB2 is the first object of this class that has been identified 
within our own Galaxy.
Young globular clusters (or young populous clusters as termed by 
\cite{hodge61}) seem not to be rare objects.
They were first identified in the LMC but are now also found using the 
{\em Hubble Space Telescope} in a variety of extragalactic star 
forming regions (Ho 1997\nocite{ho97}).
Although young globulars are often observed in association with 
starburst phenomena, they also appear to form in more quiescent 
environments, such as circumnuclear rings in relatively undisturbed 
galaxies (e.g.~Maoz et al.~1996\nocite{maoz96}).
It seems therefore plausible that such a system can also form in our 
own Galaxy, although the formation conditions are yet to be explored.

In a review about young globulars in the LMC, \cite{freeman80} asked 
why these systems ``form in the LMC and not in the Galaxy''?
Apparently, they do also form in our Galaxy, but heavy obscuration 
through the interstellar gas and dust make them difficult to detect --
or may leave them unrecognised.
Infrared surveys may help to improve the detectibility of such 
systems, although the sensitivity and confusion limits restrict such 
studies to within a few kpc.
Alternatively, one may search for the fingerprints that these systems 
leave on the surrounding interstellar medium (ISM).
An example is the thermal radio emission related to the ionisation of 
the ISM by the numerous massive stars.
Indeed, the 120 O stars in Cyg OB2 produce a Lyman continuum flux of 
roughly $10^{51}$ photons s$^{-1}$, making it probably the most important 
source of ionisation in the Cygnus X complex.
It has been suggested that the Cygnus X complex could be a large 
Str\"omgren sphere powered by Cyg OB2 (Landecker 1984\nocite{landecker84} 
and references therein), hence the search for similar complexes in 
surveys of thermal radio emission could provide hints for further young 
globulars in the Galaxy.

Another tracer may come from the emerging field of gamma-ray line 
spectroscopy.
The massive stars in Cyg OB2 are an important source of radioisotopes 
that are expelled through stellar winds and supernova explosions into 
the interstellar medium.
These isotopes eventually emit monoenergetic gamma-ray photons that 
arise from nuclear de-excitations following the radioactive decay.
An example is the 1.809 MeV gamma-ray line arising from 
the decay of radioactive $^{26}$Al, an isotope with a lifetime of 
$\sim10^6$ years (see Prantzos \& Diehl 1996\nocite{prantzos96} for a
review).
Indeed, besides the central galactic radian, the Cygnus X region is 
the most intense source of 1.809 MeV photons known in the Galaxy, 
which can be well explained by $^{26}$Al production in massive stars
within Cyg OB2 (Kn\"odlseder et al., in preparation).
A more detailed study of the galactic plane in the 1.809 MeV
gamma-ray line by the upcoming {\em INTEGRAL} mission (Vedrenne et 
al.~1999\nocite{vedrenne99}) could lead to the identification of 
further regions of high star forming activity, and hence provide 
a unique tool to unveil young globular clusters in our own Galaxy.

\begin{acknowledgements}
This publication makes use of data products from the Two Micron All Sky 
Survey, which is a joint project of the University of Massachusetts and 
the Infrared Processing and Analysis Center/California Institute of Technology, 
funded by the National Aeronautics and Space Administration and the National
Science Foundation.
The author wants to thank T. Dame for kindly providing the CO data of 
the Cygnus region.
The author is supported by an external ESA fellowship.
\end{acknowledgements}

\appendix
\section{Stellar calibrations}
\label{sec:calibrations}

\begin{table*}[th]
  \footnotesize
  \caption{\label{tab:calibration}
  Stellar calibrations employed in this work (all data are for 
  luminosity class V). 
  $\tilde{K}$ was calculated using the relation 
  $\tilde{K} = M_{K} - 0.66 \times (J-K)_{0}$ (see Appendix 
  \ref{sec:photocut}).}
  \begin{flushleft}
    \begin{tabular}{ccccccc}
    \hline
    \noalign{\smallskip}
    Sp.~Type & $M_{V}$ & $(V-K)_{0}$ & $M_{K}$ & $(J-K)_{0}$ & 
    $\tilde{K}$ & $\log M$ \\
    \hline
    \noalign{\smallskip}
    O3   & -5.78 & -0.95 & -4.83 &       &       & 2.08 \\
    O4   & -5.55 & -0.93 & -4.62 &       &       & 1.93 \\
    O4.5 & -5.44 & -0.92 & -4.52 &       & 	     &      \\
    O5   & -5.33 & -0.90 & -4.43 &       & 	     &      \\
    O5.5 & -5.22 & -0.89 & -4.33 &       &       & 1.78 \\
    O6   & -5.11 & -0.89 & -4.22 & -0.16 & -4.11 &      \\
    O6.5 & -4.99 & -0.86 & -4.13 &       &       &      \\
    O7   & -4.88 & -0.86 & -4.02 & -0.15 & -3.92 & 1.60 \\
    O7.5 & -4.77 & -0.84 & -3.93 &       &       &      \\
    O8   & -4.66 & -0.84 & -3.82 & -0.16 & -3.71 &      \\
    O8.5 & -4.55 & -0.81 & -3.74 &       &       &      \\
    O9   & -4.43 & -0.79 & -3.64 & -0.18 & -3.52 & 1.40 \\
    O9.5 & -4.32 & -0.77 & -3.55 & -0.17 & -3.44 &      \\
    B0   & -4.21 & -0.76 & -3.45 & -0.18 & -3.33 & 1.30 \\
    B0.5 & -3.51 & -0.69 & -2.82 & -0.16 & -2.71 & 1.25 \\
    B1   & -3.20 & -0.66 & -2.54 & -0.15 & -2.44 &      \\
    B1.5 & -2.80 & -0.62 & -2.18 & -0.12 & -2.10 &      \\
    B2   & -2.50 & -0.59 & -1.91 & -0.11 & -1.84 &      \\
    B3   & -1.60 & -0.49 & -1.11 & -0.10 & -1.04 & 0.88 \\
    B5   & -1.20 & -0.42 & -0.78 & -0.09 & -0.72 & 0.77 \\
    B7   & -0.60 & -0.35 & -0.25 & -0.07 & -0.20 &      \\
    B8   & -0.25 & -0.29 &  0.04 & -0.05 &  0.07 & 0.58 \\
    B9   &  0.20 & -0.17 &  0.37 & -0.02 &  0.38 &      \\
    A0   &  0.65 &  0.00 &  0.65 &  0.01 &  0.64 & 0.46 \\
    A1   &  1.00 &  0.06 &  0.94 &  0.02 &  0.93 &      \\
    A2   &  1.30 &  0.13 &  1.17 &  0.04 &  1.14 &      \\
    A3   &  1.50 &  0.20 &  1.30 &  0.05 &  1.27 &      \\
    A5   &  1.95 &  0.35 &  1.60 &  0.09 &  1.54 & 0.30 \\
    A7   &  2.20 &  0.45 &  1.75 &  0.12 &  1.67 &      \\
    A8   &  2.40 &  0.56 &  1.84 &  0.14 &  1.75 &      \\
    F0   &  2.70 &  0.79 &  1.91 &  0.20 &  1.78 & 0.20 \\
    F5   &  3.50 &  1.01 &  2.49 &  0.26 &  2.32 & 0.15 \\
    F8   &  4.00 &  1.12 &  2.88 &  0.29 &  2.69 &      \\
    G0   &  4.40 &  1.22 &  3.18 &  0.31 &  2.98 & 0.02 \\
    \noalign{\smallskip}
    \hline
    \end{tabular}
  \end{flushleft}
\end{table*}

The analysis presented in this work makes use of calibrations between 
intrinsic $K$ magnitudes and $J-K$ colours on the one side, and 
spectral types and initial masses on the other side.
The employed calibrations, valid for luminosity class V, are summarised 
in Table \ref{tab:calibration}.
The use of luminosity class V calibrations for all stars is justified 
by the young age of Cyg OB2 (Massey \& Thompson 1991\nocite{massey91}).

The relation between spectral type and absolute visual magnitudes $M_{V}$ 
has been compiled from data published by 
\cite{vacca96} for O3V - B0.5V stars,
\cite{humphreys84} for B1V - B3V stars, and
\cite{schmidt82} for B5V - G0V stars.
The intrinsic colour calibrations $(V-K)_{0}$ and $(J-K)_{0}$ were 
taken from \cite{wegner94} for spectral types O6V - B9V and from 
\cite{koorneef83} for later spectral types.
The $(V-K)_{0}$ calibration of \cite{wegner94} was extrapolated in 
the $M_{V}$,$(V-K)_{0}$ plane to spectral type O3V using the relation
\begin{equation}
 (V-K)_{0} = -0.296 + 0.114 \times M_{V}
\end{equation}
which has been obtained form a fit to the data for spectral types B2V and 
earlier.
The same relation was also used to interpolate the $(V-K)_{0}$ 
calibration for some O subtypes that were not covered by the analysis 
of \cite{wegner94}.

Based on the $M_{V}$ and $(V-K)_{0}$ calibrations, the intrinsic $K$ 
magnitudes were derived using
\begin{equation}
 M_{K} = M_{V} - (V-K)_{0} .
\end{equation}
To derive the mass-luminosity relation, the initial stellar mass to 
spectral type relation has been taken from \cite{schaerer97} 
for O and \cite{schmidt82} for later type stars.
Using a fit in the $\log M$,$M_{K}$ plane, the following relation has 
been established
\begin{equation}
 \log M = \left\{ \begin{array}{r@{\quad}l}
                   0.605 - 0.201 M_{K} + 0.0055 M_{K}^{2} & M_{K} > -3.70 \\
                   -0.630 - 0.557 M_{K} & {\rm else}
                   \end{array} \right.
\end{equation}

\section{Photometric sample selection}
\label{sec:photocut}

For the photometric sample selection two constraints have been applied 
in the $K$,$J-K$ plane.
First, all association stars with identical intrinsic magnitude $M_{K}$ 
are collected along the reddening line.
To understand the selection, recall that the apparent $K$ magnitude of a 
Cyg OB2 member star is related to the intrinsic magnitude $M_{K}$, the 
extinction $A_{K}$, and the distance modulus $DM$ via
\begin{equation}
 K = M_{K} + DM + A_{K} .
 \label{eq:rel1}
\end{equation}
The extinction of a star can be estimated from the colour excess
\begin{equation}
 E(J-K) = (J-K) - (J-K)_{0}
 \label{eq:colorexcess}
\end{equation}
using
\begin{equation}
 A_{K} = 0.66 \times E(J-K)
 \label{eq:extinction}
\end{equation}
(Rieke \& Lebofsky 1985\nocite{rieke85}).
Thus all member stars of identical intrinsic magnitude but different 
extinction lie in the $K$,$J-K$ CMD on a line of constant slope,
given by
\begin{equation}
 K = M_{K} - 0.66 \times (J-K)_{0} + DM + 0.66 \times (J-K) .
\end{equation}
Hence, selecting stars within a band
\begin{equation}
 \tilde{K}_{\rm low} \ge K - DM - 0.66 \times(J-K) \ge \tilde{K}_{\rm up}
\end{equation}
results in a reddening independent selection of stars, where
\begin{equation}
 \tilde{K} \equiv M_{K} - 0.66 \times (J-K)_{0} .
\end{equation}
The values of $\tilde{K}$ are listed in column 6 of Table \ref{tab:calibration}.
Fitting a linear relation to these data results in
\begin{equation}
 M_{K} = 0.0037 + 1.036 \times \tilde{K} .
\end{equation}

Second, the $K$ band extinction $A_{K}$ of an association star can be 
estimated by calculating its displacement along the reddening line.
To understand this, notice that for main sequence stars, $M_{K}$ is a 
function of intrinsic colour $(J-K)_{0}$, which for massive stars can be 
roughly approximated by a linear function.
From the calibration table (cf.~Table \ref{tab:calibration}) the relation
\begin{equation}
 M_{K} \approx -0.057 + 17.835 \times (J-K)_{0}
 \label{eq:rel2}
\end{equation}
has been determined for $M_{K} < 2^{\rm m}$.
Using Eqs.~(\ref{eq:colorexcess}) and (\ref{eq:extinction}) the relation 
can be rewritten as
\begin{equation}
 M_{K} \approx -0.057 + 17.835 \times \left( (J-K) - 
 \frac{A_{K}}{0.66} \right) .
\end{equation}
Replacing this approximation in Eq.~(\ref{eq:rel1}) and solving for 
$A_{K}$ result in an approximative relation between apparent magnitude 
and colour and extinction in the $K$ band:
\begin{equation}
 A_{K} \approx \frac{K + 0.057 - 17.835 \times (J-K) - DM}
                    {1 - 17.835 / 0.66} .
\end{equation}


\end{document}